# Modelling a new, low $CO_2$ emissions, hydrogen steelmaking process

A. Ranzani da Costa, D. Wagner[#] and F. Patisson[*]

*Institut Jean Lamour, Labex DAMAS, CNRS, Université de Lorraine, Nancy, France*

## Abstract

In an effort to develop breakthrough technologies that enable drastic reduction in $CO_2$ emissions from steel industry (ULCOS project), the reduction of iron ore by pure hydrogen in a direct reduction shaft furnace was investigated. After experimental and modelling studies, a 2D, axisymmetrical steady-state model called REDUCTOR was developed to simulate a counter-current moving bed reactor in which hematite pellets are reduced by pure hydrogen. This model is based on the numerical solution of the local mass, energy and momentum balances of the gas and solid species by the finite volume method. A single pellet sub-model was included in the global furnace model to simulate the successive reactions (hematite->magnetite ->wustite->iron) involved in the process, using the concept of additive reaction times. The different steps of mass transfer and possible iron sintering at the grain scale were accounted for. The kinetic parameters were derived from reduction experiments carried out in a thermobalance furnace, at different conditions, using small hematite cubes shaped from industrial pellets. Solid characterizations were also performed to further understand the microstrutural evolution. First results have shown that the use of hydrogen accelerates the reduction in comparison to CO reaction, making it possible to design a hydrogen-operated shaft reactor quite smaller than current MIDREX and HYL. Globally, the hydrogen steelmaking route based on this new process is technically and environmentally attractive. $CO_2$ emissions would be reduced by more than 80%. Its future is linked to the emergence of the hydrogen economy.

## 1 Introduction

The increase in $CO_2$ level of the troposphere, mainly due to emissions from fossil fuel combustion, is very likely the cause for the climate change observed over the past several decades, characterized by the rapid growth of global average temperatures (IPCC, 2007). Hence, to prevent global warming many countries signed the Kyoto protocol for decreasing their greenhouse gases (GHG) emissions. Among these GHG emissions, the Steel Industry is responsible for 19% of those of the industrial sector (GIEC, 2001). In this context, the main European steelmakers have launched, in 2004, a European research and development project called ULCOS ("*Ultra–Low CO₂ Steelmaking*") to reduce carbon dioxide emissions of today's best routes (1850 kg of $CO_2$/ton$_{steel}$) (Birat et al., 2008) by at least 50%. To achieve this target, breakthrough technologies for making steel were studied by ULCOS partners. In this paper, we are interested in the use of pure hydrogen ($H_2$) as reducing agent of iron ores in the direct reduction (DR) process, which could be the core process of a new, cleaner way to produce steel with lower $CO_2$ emissions, as shown below.

---

[#]Now, at ArcelorMittal, Maizières-les-Metz, France
[*] Corresponding author: Tel: +33 3 83584267 ; Fax : +33 3 83584056 ;
Email address: fabrice.patisson@univ-lorraine.fr

Indeed, in the conventional DR industrial processes (MIDREX and HYL III, 65% and 20% of world production, respectively), hematite iron ore is reduced by a gas mixture (syngas) obtained by reforming of natural gas ($CH_4$) and mainly composed of CO and $H_2$. Using pure $H_2$ as a reducing agent, only water vapour is released to the atmosphere and $CO_2$ from CO reduction is avoided, as it can be noted from the reactions involved in the reduction process:

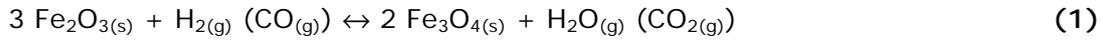

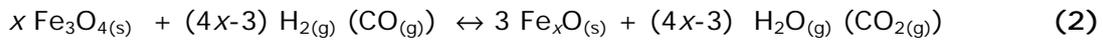

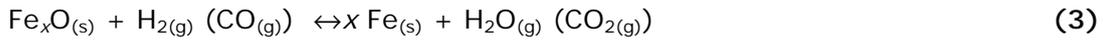

with $x$=0.95.

In the best hydrogen-based steelmaking breakthrough route studied in ULCOS (Fig.1), $H_2$ would be produced by water electrolysis using hydraulic or nuclear electricity. Iron ore would be reduced to direct reduced iron (DRI) by $H_2$ in a shaft furnace, and C-free DRI would be treated in an electric arc furnace (EAF) to produce steel. This route exhibits promising performance regarding $CO_2$ emissions: less than 300 kg $CO_2$/ton$_{steel}$, including the $CO_2$-cost of electricity (Wagner, 2008), the emissions from the DR furnace itself being almost zero. This represents an 84% cut in $CO_2$ emissions as compared to the current 1850 kg $CO_2$/ton$_{steel}$ of the best blast furnace route. This new route would thus be a more sustainable way for making steel. However, its future development is largely dependent on the emergence of a so-called $H_2$ economy, when this gas would become available in large quantities, at competitive cost, and with low $CO_2$ emissions for its production. This should be possible with a significant increase on $H_2$ demand from other industrial sectors, which could be the case of energy and transportation industries. In ULCOS, this scenario was regarded as a mid-term option. Therefore, to anticipate its possible development, it was decided to check the feasibility of using 100% $H_2$ in a Direct Reduction shaft furnace. The aim of our work was to develop a mathematical model for simulating a future DR shaft furnace operated with pure $H_2$ in order to evaluate the process.

**Fig.1** – Hydrogen-based route to steel

Most of the current DR shaft furnace models proposed in the literature deal with the reduction by the gas mixture $H_2$/CO; and they also greatly simplify the description of physical-chemical and thermal phenomena involved. Yu and Gillis (1981), Takenaka et al. (1986) and Negri et al. (1995) considered in their models one-directional (vertical) fluxes of solid and gas species inside the reactor. However, in current industrial DR furnaces, the reducing gas is mainly fed radially, through holes in the side wall of the lower part of the shaft, which makes gas flow two-dimensional. Concerning the kinetic mechanisms which control the reduction of iron ore, the most common simplification is to adopt the classical unreacted shrinking core model (USCM) to describe the reaction. Kam and Hughes (1981) and Parisi and Laborde (2004) considered this kinetic model to describe one overall reaction of hematite to metallic iron, by $H_2$ and CO, without taking into account the intermediary oxides ($Fe_2O_3 \rightarrow Fe_3O_4 \rightarrow Fe_xO \rightarrow Fe$). Using the same kinetic approach, Tsay et al. (1976) described the three intermediary reactions in their 1D bed reactor model.

Our present numerical model is based on the mathematical description of the detailed physical, chemical and thermal phenomena occurring in a shaft furnace operated with pure $H_2$. It is a 2D, axisymmetrical and steady-state model, written in FORTRAN 90, using the finite volume method and based on the numerical solution of the local mass, energy and momentum balances of the gas and solid species. It describes the three reduction reactions ($Fe_2O_3 \rightarrow Fe_3O_4 \rightarrow Fe_xO \rightarrow Fe$) involved in the process. A singlepellet sub-model is included in the shaft furnace model to simulate the successive reactions, the different steps of mass transfer and possible iron sintering at the particle scale. The kinetic parameters of the model were derived from experiments (thermogravimetry and sample characterization).

## 2 Experimental work

Although numerous studies were devoted to one or another aspect of the kinetics of iron ore reduction by CO or $H_2$, neither general kinetic model nor unique kinetic parameter values can be directly used for predicting the reduction rate in variable conditions of gas composition and temperature. Not only do reported rate constants differ by orders of magnitude, but the activation energies also vary over a wide range. For example, the activation energy for the wustite-iron reduction by $H_2$ varies from 200kJ/mol (Gaballah et al, 1972), to 117kJ/mol (Takenaka et al., 1986) and 92kJ/mol (Tsay et al., 1976). Inconsistencies can also be found in works of the same authors (Valipour et al., 2006; Valipour and Khoshandam, 2009). These discrepancies may be attributed to the wide diversity of experimental conditions (reduction temperature and pressure, gas flow and gas composition) and to the starting material (mineralogical composition, crystal size, porosity and pore distribution of the ore).

To clarify the kinetic mechanisms of the reduction of hematite pellets by pure $H_2$, we used thermogravimetry (weight-loss technique) for accurate determination and continuous recording of the weight loss of iron ore during the reduction process as a function of time. The experiments were carried out using a SETARAM TAG 24 thermobalance, which features a pair of symmetric furnaces. This arrangement enables high precision on mass variation measurements (μg accuracy) since it eliminates buoyancy and drag forces effects. A specific steam generator was coupled to the furnace to possibly add controlled water content to the reaction gas. Not to exceed the maximum weight loss imposed by the thermobalance (400 mg), we shaped small hematite cubes (3.5-5 mm side, 200-550 mg weight) from industrial pellets (4g weight), as thermogravimetric samples. Pellets used were Brazilian CVRD hematite pellets composed of approximately 96% of $Fe_2O_3$ and 4% of other oxides (CaO, $SiO_2$, $Al_2O_3$, MgO, MnO, etc.). The tests were run under isothermal conditions. A single $Fe_2O_3$ cube wrapped in a Pt wire was directly hung to the balance beam, placed inside the furnace (18 mm of inner diameter) and preheated to a chosen temperature under an inert gas atmosphere of He. When the desired temperature was reached and stabilized, $H_2$ was introduced into the furnace and maintained until the end of the reaction, when the weight loss of the sample was no more significant. The reduced sample was then cooled to room temperature under an inert He atmosphere. Experiments were made at temperatures between 500 and 990°C. Gas compositions were 100 % $H_2$ and 60 vol. % $H_2$ in He. In some cases, up to 4 vol. % $H_2O$ were added to study the effect of water in the reducing gas. Experiments showed that a gas flow of 200mL/min was sufficient to ensure a good external mass transfer and not affect the reduction curves. In addition to normal complete experiments, partially reduced samples were prepared by interrupting the $H_2$ flow inside the furnace before complete reduction took place. The complete and partially reduced samples were analyzed by X-ray diffraction (XRD), scanning electron microscopy (SEM) and Mössbauer spectrometry to characterize the morphological evolution during the reduction and also identify the mechanisms that control the global reaction rate.

## 3   Experimental results and discussion

### 3.1   Influence of the temperature

The reduction curves obtained experimentally by thermogravimetry at different temperatures are given in Fig.2. The conversion plotted on y-axis represents the fractional oxygen lost by the iron ore in the course of the reduction from hematite to metallic iron (hematite/magnetite/wustite/iron).

**Fig.2**. Influence of the temperature on the reduction kinetics of small hematite cubes (5mm side, 550 mg weight) by $H_2$ (200 mL/min of total gas flow, $H_2$/He 60/40 %vol.)

These curves show that 800°C seems to be the optimum temperature to reduce hematite cubes by $H_2$ in lesser time (Fig.2). For experiment temperatures under 800 °C (Fig.2, left), the higher the temperature, the faster the reaction, except at 700 °C, where the reaction slows down at high conversions. At 850°C, 900°C and 990°C, (Fig.2, right), even if an increase in temperature first accelerates the reduction, a significant slowing down of the rate at the end of the reaction increases the total time of reduction. This slowing down of the rate starts earlier and earlier as the temperature increases. In order to clarify this effect of the temperature on the reduction rate, cubes completely reduced at different temperatures were observed by SEM (Fig.3).

**Fig.3.** Influence of the temperature on the morphology of the iron obtained after reduction of hematite cubes (5mm side, 550 mg weight) by $H_2$ (200 mL/min of total gas flow, $H_2$/He 60/40 %vol.).

It can be seen from Fig.3 that the higher the temperature, noticeably denser is the iron formed. The iron obtained at high reduction temperatures (990°C) presents fewer, bigger and isolated pores, with a smoother surface as compared with the iron obtained at 600°C. We attributed this densification to the tendency of the freshly formed iron to sinter. Sintering is a mass transfer phenomenon, activated by the temperature and time-dependent, which decreases the specific surface area of the material. Its consequences are a decrease in pore volume, variation of the pore geometry and growth of the grains. At high temperatures, which promote sintering, the pores get thinner and eventually disappear, causing a densification of the iron obtained, as observed in Fig.3. This makes gas phase diffusion through the iron pores more difficult. Therefore, solid phase diffusion of oxygen, a slower process, is the only possible mass transfer mechanism to reach the wustite that remained entrapped inside this dense iron layer and to complete the reaction. This could explain the slowing down of the rate observed at high temperatures, where sintering is favoured, and at high conversions, when a significant quantity of iron phase is already present in the sample.

### 3.2 Microstructure evolution during the reduction: Interrupted experiments.

With the aim of understanding the morphological evolution of the samples during the reduction, some experiments carried out at 800°C were interrupted before completion. Partially-reduced samples were observed by SEM (Fig.4), and XRD and Mössbauer spectrometry were used to quantify the intermediary oxides present in the samples.

The initial hematite sample is made up of relatively large, dense grains, with angular sides (Fig.4-a). After a few seconds, the surface, which is mainly magnetite and wustite, becomes covered with small pores (Fig.4-b) but the main structure of the initial sample remains almost unchanged. Pores slightly enlarge when wustite appears (Fig.4-c), but no significant change is observed in the microstructure. Around 50% of conversion, when the iron content in the sample is quite significant (66% Fe and 34% FeO) (Fig.4-d), the initial grainy microstructure breaks up into smaller particles, which we termed "crystallites", and is progressively replaced by a molten-like structure, characteristic of the sponge iron (Fig.4-e).

One of the partially reduced cubes (experiment at 800°C, interrupted at the stage of wustite) was impregnated with resin and polished. A cross-section of this sample observed by SEM (Fig.4-f) further reveals the structure of wustite: very porous and sub-divided into smaller grains, the crystallites.

**Fig.4.** Microstructure evolution of the solids during reduction at 800°C by $H_2$ (200 mL/min of total gas flow, $H_2$/He 60/40 %vol.). Interrupted experiments: (a) initial; (b): after 89s; (c): after 206s; (d): after 400s; (e): fully reduced; (f): after 400s, cross-section of polished sample

## 4 Formulation of the mathematical model

### 4.1 Principle of the mathematical model

We developed a mathematical model of a shaft furnace for the reduction of iron ore by hydrogen and its corresponding numerical code, called REDUCTOR. Our objective was to build a valuable simulation tool that would help us to optimize the operating conditions and the design of a future, clean, direct-reduction reactor operated with pure hydrogen.

In an industrial DR process, like MIDREX or HYL, iron oxide, in the form of pellets or lump ore, is introduced at the top of the shaft furnace, through a hopper, descends by gravity and encounters a counter-current flow of syngas, a mixture of mostly CO and $H_2$, produced by reforming of natural gas. This reducing gas heats up the descending solids and reacts with the iron oxide, converting it into metallic iron (DRI) at the bottom of the cylindrical upper section of the reactor (i.e., the reduction zone). $CO_2$ and $H_2O$ are released. For production of cold DRI, the reduced iron is cooled and carburized by counterflowing cooling gases in the lower conic section of the furnace (cooling zone). The DRI can also be discharged hot and fed to a briquetting machine for production of HBI (Hot Briquetted Iron), or fed hot directly to an EAF (Electric Arc Furnace).

The REDUCTOR model was developed to simulate the reduction zone of a shaft furnace, similar to a MIDREX one. Thus, a counter-current moving bed cylindrical reactor, with 9 m height and 6.6 m diameter is described (Fig.5). The iron ore is fed at the top of the furnace as pellets. The gas is mainly injected laterally into the bed, through a 30-cm high ring-like inlet situated at 20 cm above the bottom of the reactor. A small part (2%) of the gas flow is injected from the bottom of the furnace. Reducing gas is composed of $H_2$, $H_2O$ an $N_2$. For the sake of simplicity, we consider that the pellets are spherical (1.2-cm diameter) and composed of 100% of hematite.

The mathematical model itself is two-dimensional, axisymmetrical and steady-state. It is based on the numerical resolution of local mass, energy and momentum balances using the finite volume method (Patankar, 1980), and on a single-pellet sub-model for calculating the reduction kinetics. Calculated results include all of the relevant variables of the process (local solid and gas temperatures, compositions and velocities, reactions rates, conversion, etc.).

**Fig. 5.** Principle of the REDUCTOR model

*4.2 Kinetic sub-model of a single pellet*

A kinetic sub-model was built according to the experimental findings to simulate the reduction of a single pellet by $H_2$. This model is used as a subroutine of the shaft furnace model and it predicts the reaction rate as a function of the local reduction conditions (temperature and gas composition) inside the reactor. It is based on the law of additive reaction times (Sohn, 1978). This law states that the time required to attain a certain conversion is approximately the sum of characteristic times ($\tau_i$) corresponding to each elementary step of mass transfer. In our case, these steps are: the chemical reactions, $H_2$ and $H_2O$ gas diffusion through the pores of the pellets, oxygen solid diffusion through the dense iron layer formed and $H_2$ and $H_2O$ mass transfer through the boundary layer surrounding the pellet. The characteristic time of step *i*, $\tau_i$, is the time necessary to attain complete conversion in the case of a system controlled only by this elementary step *i*. This is a useful approximation in the case of gas-solid systems, where the different kind of mass transfer can be considered as a series of resistances. Its great advantage is its ability to represent intermediate (mixed) kinetic regimes in a closed-form equation, drastically reducing the computation time, particularly in the case of complex reactor models (Patisson et al., 2006).

Based on the SEM images of partially reduced samples (Fig.4), assumptions were made concerning the morphology of the pellets during the reduction. A pellet at hematite and magnetite stagesis supposed to be an agglomerate of dense spherical grains of the same diameter (25μm), separated by inter-grain porosity (0.1). At the stage of wustite, however, the grains of the pellet break up into crystallites (Fig.4-f) and become porous. Thus, they can be considered as a combination of dense spherical crystallites (2μm diameter), and pores (intra-grain porosity 0.53). As the wustite crystallites reduce to metallic iron, they become themselves porous due to molar volume difference between wustite and iron. Therefore, besides the inter-grain porosity, we attribute also intra-grain and intra-crystallite porosities to the pellets at wustite stage.

In the case of hematite to iron transformation, three reactions (Eq.1 to 3) are involved, each one with a characteristic time $\tau_{ch,i}$. Moreover, considering the different porosity levels inside the pellet, gas diffusion must be taken into account through inter-grain, intra-grain and intra-crystallite pores, the latter appearing only at wustite stage. In the case of intra-grain and intra-crystallite pores, the Knudsen-type diffusion is not negligible compared to molecular diffusion. Oxygen solid diffusion through the dense iron layer formed around the crystallites is also considered at high temperatures and conversions (wustite-iron reaction). The densification of the iron is also described as a function of temperature and time by introducing a characteristic time for sintering ($\tau_{sint}$). Detailed equations to calculate the characteristic times ($\tau_i$) and reaction rates ($r_i$) are given in the work of Wagner (2008) and summed up in Table 1. Notation is given at the end of the paper.

| | External transfer | Inter-granular diffusion | Intra-granular diffusion | Intra-crystallite diffusion with sintering | Chemical reaction |
|---|---|---|---|---|---|
| | | | Global rate | | |
| (1) Hematite → Magnetite | $\tau_{ext_1} = \dfrac{c_{Fe_2O_3,in,s}\, d_p}{18 k_{ext} c_t \left( x_{H_2} - x_{H_2eq} \right)}$ | $\tau_{dif_1} = \dfrac{c_{Fe_2O_3,in,s}\, d_p^2}{72 D_{H_2-eff,g} c_t \left( x_{H_2} - x_{H_2eq} \right)}$ | | | $\tau_{r_1} = \dfrac{c_{Fe_2O_3,in,s}\, d_{gr}}{6 k_{s_1} c_t \left( x_{H_2} - \dfrac{x_{H_2eq}}{K_1} \right)}$ |
| | $r_1 = \frac{1}{3} c_{Fe_2O_3,in,s} \left\{ \tau_{ext_1} + 2\tau_{dif_1}\left[ (1-X_1)^{-1/3} - 1 \right] + \dfrac{\tau_{r_1}}{3}(1-X_1)^{-2/3} \right\}^{-1}$ | | | | |
| (2) Magnetite → Wüstite | $\tau_{ext_2} = \dfrac{8 c_{Fe_3O_4,in,s}\, d_p}{57 k_{ext} c_t \left( x_{H_2} - x_{H_2eq} \right)}$ | $\tau_{dif_2} = \dfrac{2 c_{Fe_3O_4,in,s}\, d_p^2}{57 D_{H_2-eff,g} c_t \left( x_{H_2} - x_{H_2eq} \right)}$ | | | $\tau_{r_2} = \dfrac{c_{Fe_3O_4,in,s}\, d_{gr}}{2 k_{s_2} c_t \left( x_{H_2} - \dfrac{x_{H_2eq}}{K_2} \right)}$ |
| | $r_2 = c_{Fe_3O_4,in,s} \left\{ \tau_{ext_2} + 2\tau_{dif_2}\left[ (1-X_2)^{-1/3} - 1 \right] + \dfrac{\tau_{r_2}}{3}(1-X_2)^{-2/3} \right\}^{-1}$ | | | | |
| (3) Wüstite → Iron | $\tau_{ext_3} = \dfrac{c_{Fe_{0.95}O,in,s}\, d_p}{6 k_{ext} c_t \left( x_{H_2} - x_{H_2eq} \right)}$ | $\tau_{dif,p} = \dfrac{c_{Fe_{0.95}O,in,s}\, d_p^2}{24 D_{H_2-eff,p} c_t \left( x_{H_2} - x_{H_2eq} \right)}$ | $\tau_{dif,gr} = \dfrac{c_{Fe_{0.95}O,in,s}\, d_{gr}^2}{24 D_{H_2-eff,gr} c_t \left( x_{H_2} - x_{H_2eq} \right)}$ | $\tau_{dif,cr}(t) = \dfrac{c_{Fe_{0.95}O,in,s}\, d_{cr}^2}{24 D_{gr,cr}(t) c_t \left( x_{H_2} - x_{H_2eq} \right)}$ | $\tau_{r_3} = \dfrac{c_{Fe_{0.95}O,in,s}\, d_{cr}}{2 k_{s_3} c_t \left( x_{H_2} - \dfrac{x_{H_2eq}}{K_3} \right)}$ |
| | $r_3 = c_{Fe_{0.95}O,in,s} \left\{ \tau_{ext_3} + 2\left( \tau_{dif,p} + \tau_{dif,gr} + \tau_{dif,cr} \right)\left[ (1-X_3)^{-1/3} - 1 \right] + \dfrac{\tau_{r_3}}{3}(1-X_3)^{-2/3} \right\}^{-1}$ | | | | |

**Table 1.** Expressions of the characteristic times and reaction rates.

As shown by Fig.6-left, this kinetic model represents fairly well the experimental curves obtained with the reduction of hematite cubes with hydrogen. By plotting the variation of the mass fractions of each oxide with time (Fig.6-right), it appears that the first two reactions are very fast and that the wustite-iron reduction controls the overall transformation. One can also notice, on these curves at 900°C, the effect of sintering around 80% of conversion, when the reaction rate slows down.

**Fig.6.** Simulation results from the kinetic model describing the $H_2$ reduction of a single pellet at 900°C (200 mL/min of total gas flow $H_2$/He 60/40 %vol.): Left - comparison with thermogravimetric experiments; Right – evolution of the mass fractions of each oxide with time

It is important to emphasize that kinetic parameters used in this kinetic model of a single pellet, as well as in the multiparticle reactor model described below, were obtained from the experimental tests carried out with the small cubes. First, the extrapolation to an entire pellet was made changing only the size and shape of the particle. This seems a reasonable approximation since cubes and pellets are made of the same raw material; anyway, further thermogravimetric experiments should be undertaken with industrial pellets to confirm this hypothesis. Second, from a pellet in the thermobalance to the pellets in the multiparticle reactor, only the external transfer, between the bulk gas and the pellet outer surface, changes. The external transfer coefficients $k_{ext}$ are thus calculated from different correlations in both cases (Culham et al., 2001, for a cube in the thermobalance, Wakao and Kaguei, 1982, for a pellet in the multiparticlebed ).

*4.3 The Main equations of the REDUCTOR model*

The main equations of the REDUCTOR model are local mass, energy and momentum balances. While considering a 2D flow for the gas and 1D flow for the solid within a cylindrical coordinate system, the following assumptions were made: steady-state, axisymmetry, heat of reaction released in the solid phase. The main equations of the model are presented in (4-14). Where are the equations? Do the only appear in the PDF version?

*Mass balance for gaseous species i (mol m$^{-3}$ s$^{-1}$)*

$$\frac{1}{r}\frac{\partial\left( r c_t x_i u_{g,r} \right)}{\partial r} + \frac{\partial\left( c_t x_i u_{g,z} \right)}{\partial z} = \frac{1}{r}\frac{\partial}{\partial r}\left( r c_t D_r \frac{\partial x_i}{\partial r} \right) + \frac{\partial}{\partial z}\left( c_t D_z \frac{\partial x_i}{\partial z} \right) + S_i \qquad \textbf{(4)}$$

with $\quad -S_{H_2} = S_{H_2O} = r_1 + \dfrac{16}{19}r_2 + r_3$ (5)

*Mass balance for solid species j (kg m$^{-3}$ s$^{-1}$)*

$$-\frac{\partial\left(\rho_b u_{s,z} w_j\right)}{\partial z} = S_j \tag{6}$$

with $\quad S_{Fe_2O_3} = -3M_{Fe_2O_3}r_1$ (7) $\qquad S_{Fe_3O_4} = M_{Fe_3O_4}\left(2r_1 - r_2\right)$ (8)

$\qquad S_{Fe,O} = M_{Fe,O}\left(\dfrac{60}{19}r_2 - r_3\right)$ (9) $\qquad S_{Fe} = 0.95 M_{Fe} r_3$ (10)

*Heat balance for the gas (W m$^{-3}$)*

$$\rho_g c_{pg}\left(u_{gr}\frac{\partial T_g}{\partial r} + u_{gz}\frac{\partial T_g}{\partial z}\right) = \frac{1}{r}\frac{\partial}{\partial r}\left(r\lambda_g\frac{\partial T_g}{\partial r}\right) + \frac{\partial}{\partial z}\left(\lambda_g\frac{\partial T_g}{\partial z}\right) + a_b h\left(T_s - T_g\right) +$$
$$+\left(r_1 + \frac{16}{19}r_2 + r_3\right)\int_{T_g}^{T_s}\left(c_{pH_2O} - c_{pH_2}\right)dT \tag{11}$$

*Heat balance for the solid (W m$^{-3}$)*

$$-\rho_b u_{s,z} c_{ps}\frac{\partial T_s}{\partial z} = \frac{1}{r}\frac{\partial}{\partial r}\left(r\lambda_{eff,r}\frac{\partial T_s}{\partial r}\right) + \frac{\partial}{\partial z}\left(\lambda_{eff,z}\frac{\partial T_s}{\partial z}\right) + a_b h\left(T_g - T_s\right) + \sum_{n=1,2,3}\left(-r_n\Delta_r H_n\right)$$

$$\tag{12}$$

*Momentum equation combined with gas continuity equation*

$$\frac{1}{r}\frac{\partial}{\partial r}\left(r\frac{c_t}{K}\frac{\partial p}{\partial r}\right) + \frac{\partial}{\partial z}\left(\frac{c_t}{K}\frac{\partial p}{\partial z}\right) = 0 \qquad (mol\ m^{-3}\ s^{-1}) \tag{13}$$

with: $\quad K = 150\dfrac{\left(1-\varepsilon_b\right)^2}{\varepsilon_b^3 d_p^2}\mu_g + 1.75\dfrac{1-\varepsilon}{\varepsilon_b^3 d_p}\rho_g u_g \quad (kg\ m^{-3}\ s^{-1})$ (14)

*Boundary conditions*

– solid inlet at the top – Dirichlet conditions for the solid characteristics $T_s$, $w_{Fe_2O_3}$, $w_{Fe_3O_4}$, $w_{Fe,O}$, $w_{Fe}$, $u_{s,z}$, which are known, and for the outlet gas, known pressure $p_{out}$ and negligible axial diffusion fluxes compared to the convective ones, i.e. $\dfrac{\partial T_g}{\partial z} = \dfrac{\partial x_i}{\partial z} \approx 0$,

– side wall gas inlet – Dirichlet conditions: $T_g$, $x_{H_2}$, $x_{H_2O}$, $x_{N_2}$, $u_{g,r}$ are known,

– bottom gas inlet – Dirichlet conditions for the same gas characteristics $T_g$, $x_{H_2}$, $x_{H_2O}$, $x_{N_2}$, $u_{g,z}$, which are known, and negligible axial solid conduction flux compared to the convective one, i.e. $\dfrac{\partial T_s}{\partial z} \approx 0$ ,

– symmetry axis – Zero fluxes: $\dfrac{\partial T_s}{\partial r} = \dfrac{\partial T_g}{\partial r} = \dfrac{\partial x_i}{\partial r} = 0$ ,

– side wall (except gas inlet) – Zero fluxes: $\dfrac{\partial T_s}{\partial r} = \dfrac{\partial T_g}{\partial r} = \dfrac{\partial x_i}{\partial r} = 0$ .

*4.4 Main simulation results: parametric study*

First of all, a reference case was simulated. As mentioned above (*section 4.1*), the geometry of the problem is that of the reduction zone of a conventional MIDREXshaft furnace. The dimensions of the reactor and the known characteristics of the inlet gas and solid streams are indicated in Fig.7-a. The inlet solid flow rate used in this reference case corresponds to an annual production of 1Mton of iron. To keep the driving force for the reduction high and, above all, to bring the necessary heat for the reduction, 3.8 times the stoichiometric gas flow is injected inside the bed. Hematite pellets ($d_p$=12 mm) are fed at the top of the furnace at 25$^o$C, while gas (H$_2$/H$_2$O, 98/2 %vol.) is injected at 800$^o$C. Fig.7-b to Fig.7-e show the evolution of solid mass fractions throughout the reactor, at these reference conditions. As expected, the first two reactions are very fast compared withthe wustite-iron transformation. It should be pointed out that complete conversion to metallic iron is attained 2m above the bottom of the furnace (solid outlet). Thus, in these conditions, a 4-m high reactor would suffice to achieve the reduction. This is an important result because, in a conventional MIDREX process, a 9-m high reduction zone is needed to attain 92% of conversion at the solid outlet, using syngas as reducing agent in the same conditions.

**Fig.7**. Reference case: (a) operating parameters; (b), (c), (d), (e): calculated mass fraction of solid species inside the reactor.

Other simulations were carried out to test the influence of some operating conditions and physical-chemical parameters. Plots in Fig.8 show the iron mass fractions calculated in different conditions and are to be compared with Fig. 7-e. Fig.8-a and Fig.8-b illustrate the influence of the gas inlet temperature on the iron mass fraction inside the furnace. When the reducing gas is injected at 600$^o$C (Fig.8-a), the solid leaves the reactor with a low mean metallic iron content and the conversion is not uniform along the radial axis (2D effect). The higher degrees of reduction can be found only near the lateral gas inlet. In a large zone of the bed, the gas temperature is not high enough to promote the reduction. On the other hand, when the gas inlet temperature is 1000$^o$C (Fig.8-b), the conversion is neither complete. Comparing with the reference case where gas is injected at 800$^o$C (Fig.7-e), the reduction at 1000$^o$C is noticeably slower. This is in agreement with the kinetic study (800$^o$C was found to be the optimum reduction temperature for hematite cubes) and results from the occurrence of sintering at high temperatures (here, 1000 °C).

**Fig.8.**Iron mass fraction inside the reactor: (a)T$_g$=600°C; (b)T$_g$= 1000°C; (c) d$_p$=6mm; (d) d$_p$=24mm

Fig.8-c and 8-drender the influence of pellet size on the rate of reduction. The complete reduction of 6-mm diameter pellets takes place in the first 2 m of the reactor height (Fig.8-c), while the simulation of the reference case showed that 4 m were necessary to complete the reduction of 12-mm diameter pellets (Fig.7-e). Conversely, with larger pellets ($d_p$=24mm) (Fig.8-d), it is only possible to obtain a mean metallic iron content of 75% at the bottom of the reactor. These results express that the diffusion inside the pellets is one of the controlling mechanisms of the hematite-iron transformation and also reveal that decreasing the pellet size could represent an interesting option to accelerate the reaction.

## 5 Conclusions

The process of direct reduction of iron ore in a shaft furnace operated with pure $H_2$was evaluated as a promising mid-term breakthrough technology to produce steel with a dramatic (more than 80%) reduction in $CO_2$ emissions as compared to the current blast furnace route. We developed a two-dimensional, steady-state model of this future process to evaluate *a priori* its performance. This model is based on the numerical solution of local balance equations using the finite volume method. A kinetic sub-model, built using the concept of additive reaction times, was incorporated in the furnace model to simulate the successive reactions involved in the reduction of a singlepelletby pure $H_2$. The kinetic laws were derived from thermogravimetric experiments performed with small hematite cubes shaped from industrial pellets. An original feature of this model is the description of iron densification by sintering, at temperatures higher than 800$^o$C. The kinetic parameters obtained for small cubes were extrapolated to full-size pellets considering similar kinetic behaviour. However, further experiments should be performed with industrial pellets to confirm this assumption and to validate the kinetic model for full-size pellets.

The first results from the model showed that complete conversion to metallic iron using pure $H_2$ could be obtained using a more compact reactor than current industrial DR furnaces, which usesyngas (CO+$H_2$) as a reducing agent. This confirms the fact that reduction by $H_2$ is faster than that by CO. A parametric study showed that the size of the pellets and the temperature of the inlet gas have a strong influence on the reduction rate: the smaller the pellet diameter, the faster the reduction and thus the more compact the reactor. The optimum temperature was found to be 800$^o$C. At higher temperatures, the densification of iron due to sintering causes the reaction to slow down at high conversions. The next steps of this work should be to verify if the kinetics used are valid for entire pellets and to adapt the model for the reduction of iron ore by $H_2$/CO mixtures to validate the model against operation data of existing industrial DR process.

Finally, the results obtained so far show the technical feasibility and the environmental interest of the hydrogen-based steelmaking route. If the so-called hydrogen economy would emerge, this new hydrogen steelmaking route would become a cleaner, more sustainable way for making steel.

## Notation

$a_b$= specific surface area of the bed($m^2/m^3$)

$c$ = molar concentration (mol.m$^{-3}$)

$c_{pg}$ = molar specific heat of the gas (J.mol$^{-1}$.K$^{-1}$)

$c_{ps}$ = mass specific heat of the solid (J.kg$^{-1}$.K$^{-1}$)

$d$ = diameter (m)

$D$ = diffusion or dispersion ($D_a$,$D_r$) coefficients (m$^2$/s)

$h$ = heat transfer coefficient (W.m$^{-2}$.K$^{-1}$)

$K$ = permeability coefficient (kg.m$^{-3}$.s$^{-1}$)

$k$ = mass transfer coefficient, or reaction rate constant (m.s$^{-1}$)

$M$ = molar weight (kg.mol$^{-3}$)

$p$ = gaspressure (Pa)

$r$ = radius (m)

$r$ = reaction rate (mol.m$^{-3}$.s$^{-1}$)

$S$ = source term (mol.m$^{-3}$.s$^{-1}$)

$T$ = temperature (K)

$u$ = velocity (m.s$^{-1}$)

$w_j$ = mass fraction of $j$ in solid

$X$ = degree of conversion

$x_i$ = molar fraction of $i$ in the gas

$z$ = height (m)

*Greek:*

$\Delta_r H_n$ = heat of reaction $n$ ($n$ = 1, 2 or 3) (J.mol$^{-1}$)

$\varepsilon$ = porosity

$\lambda$ = thermal conductivity (W.m$^{-1}$.K$^{-1}$)

$\mu_g$ = viscosity of the gas (Pa.s)

$\rho_g$ = mass density of the gas (kg.m$^{-3}$)

$\rho_b$ = apparent mass density of the bed (kg.m$_{bed}^{-3}$)

*Subscripts:*

$\infty$ = bulk, i.e. in the external gas far from the pellet

1 = reaction hematite➜magnetite, (1) in the text

2 = reaction magnetite➜wustite, (2) in the text

3 = reaction wustite➜iron, (3) in the text

$b$ = bed

$ch$ = chemical

$cr$ = crystallite

$dif$ = diffusion

$eff$ = effective

$eq$ = equilibrium

$ext$ = external

$g$ = gas

$gr$ = grain

$i, j$ = species

$in$ = initial (in particles) or inlet (of the descending bed)

$p$ = pellet

$r$ = radial

$s$ = solid

$t$ = total (all species of the phase)

$x$ = sub-stoichiometry of iron in wustite = 0.95

$z$ = axial

## Acknowledgements

This piece of work was financed by the European Commission within the 6$^{th}$ framework programme, project n° 515960: ULCOS (Ultra low $CO_2$ steelmaking). We warmly thank Dr. J.P. Birat, the coordinator of ULCOS for his continuous support.

# References

Birat, J.P., Borlée, J., Korthas, B., van der Stel, J., Meijer, K., Günther, C., Halin, M., Bürgler, T., Lavelaine, H., Treadgold, C., Millar, I., Sert, D., Torp, T., Patisson, F., Paya, B., Burstrom, E., Breakthrough solutions to the $CO_2$ challenge explored by the European steel industry: the ULCOS program. A progress report in the Spring of 2008, SCANMET III Conference, 3[rd] International Conference on Process Development in Iron and Steelmaking, 8-11 June 2008, Luleå, Sweden, Proceedings edited by SwereaMefos, Luleå, 1-15.

Culham, J.R.,Yovanovich, M.M.,Teerlstra, P., Wang, C.S., Refai-Ahmed, G.,Tain, R.M., 2001, Simplified analytical models for forced convection heat transfer from cuboids of arbitrary shape, Journal of Electronic Packaging, 123, 3, 182-188.

Gaballah, I., Bert, P., Dufour, L.C., Gleitzer, C., 1972, Etude cinétique de la réduction de la wüstite par l'hydrogène et les mélanges $CO+H_2$ ;Mise en évidence de trichites, MémoiresScientifiques Rev. Métallurg.,LXIX, N° 7-8, 523-530

GIEC, 2001, "Changementsclimatiques 2001", rapport de synthèse du GIEC (IPCC), Genève

IPCC, 2007,Fourth Assessment Report, Working Group I - The Physical Science Basis, Chapter 7, 501

Kam, E., K., T., Hughes, R., 1981, A model for the direct reduction of iron ore by mixtures of hydrogen and carbon monoxide in a moving bed, Trans.IChemE, 59, 196-206

Negri, E.D., Alfano, O.M. and Chiovetta, M.G., 1995, Moving-Bed Reactor Model for the Direct Reduction of Hematite. Parametric Study, Industrial & Engineering Chemistry Research, 34, (12), 4266-4276

Parisi, D. R. and Laborde, M. A., 2004, Modelling of counter current moving bed gas-solid reactor used in direct reduction of iron ore, Chemical Engineering Journal, 104, (1-3), 35-43

Patankar, S.V., 1980, Numerical heat transfer and fluid flow, New York, Hemisphere Publishing Corp.

Patisson, F., Dussoubs B., Ablitzer D., Using Sohn's law of additive reaction times for modeling a multiparticle reactor. The case of the moving bed furnace converting uranium trioxide into tetrafluoride" Sohn International Symposium "Advanced processing of metals and materials, 27-31 August 2006, San Diego. Proceedings edited by F. Kongoli and R.G. Reddy, TMS, vol. 1 "Thermo and physicochemical principles: non-ferrous high-temperature processing", 141-153. http://hal.archives-ouvertes.fr/hal-00265646/fr/

Sohn, H. Y., 1978, The law of additive reaction times in fluid-solid reactions, Metallurgical Transactions B: Process Metallurgy, 9B, (1), 89-96

Takenaka, Y., Kimura, Y., Narita, K., Kaneko, D., 1986, Mathematical model of direct reduction shaft furnace and its application to actual operations of a model plant, Computers and Chemical Engineering, 10, (1), 67-75

Tsay, O. T .,Ray, W. H., J.Szekely, 1976, Modelling of hematite reduction with hydrogen plus carbon monoxide mixtures – Part II – The direct reduction process in a shaft furnace arrangement, AIChE Journal, 22, (6), 1072-1079

Valipour, M. S., Khoshandam, B., 2009, Numerical modelling of non-isothermal reduction of porous wustite pellet with syngas, Ironmaking and Steelmaking, 36, (2), 91-96

Valipour, M. S., MotamedHashemi, M. Y., Saboohi, Y., 2006, Mathematical modeling of the reaction in an iron ore pellet using a mixture of hydrogen, water vapor, carbon monoxide and carbon dioxide: An isothermal study, Advanced Powder Technology, 17, (3), 277-295.

Wagner, D., Étude expérimentale et modélisation de la réduction du minerai de fer par l'hydrogène. PhD thesis, Nancy-Université, France, 2008    http://tel.archives-ouvertes.fr/tel-00280689/fr/

Wakao, N., Kaguei, S.,1982, Heat and mass transfer in packed beds, Gordon and Breach science publishers.

Yu, K.O., Gillis, P.P., 1981, Mathematical simulation of direct reduction, Metallurgical Transactions B, 12B, 111-120

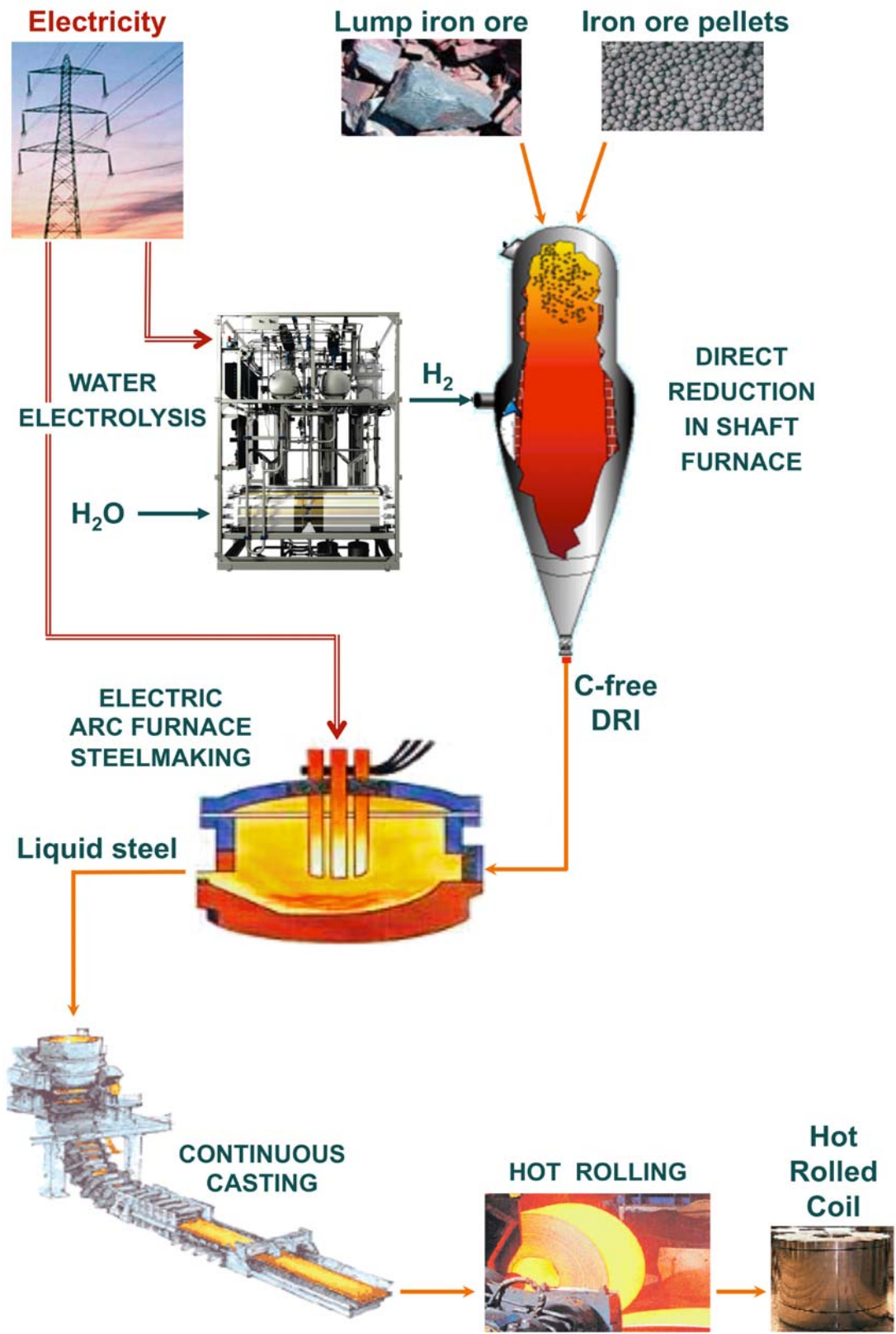

**Electricity**

**Lump iron ore**

**Iron ore pellets**

WATER
ELECTROLYSIS

$H_2$

DIRECT
REDUCTION
IN SHAFT
FURNACE

$H_2O$

C-free
DRI

ELECTRIC
ARC FURNACE
STEELMAKING

Liquid steel

CONTINUOUS
CASTING

HOT ROLLING

Hot
Rolled
Coil

**Fig. 1** – Hydrogen-based route to steel

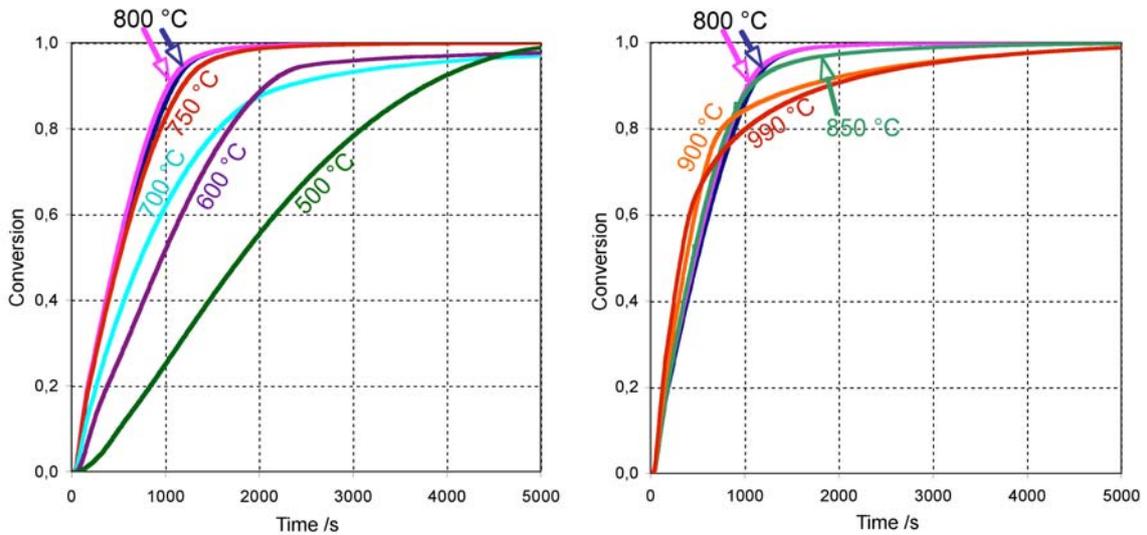

**Fig. 2.** Influence of the temperature on the reduction kinetics of small hematite cubes (5 mm side, 550 mg weight) by $H_2$ (200 mL/min of total gas flow, $H_2$/He 60/40 %vol.)

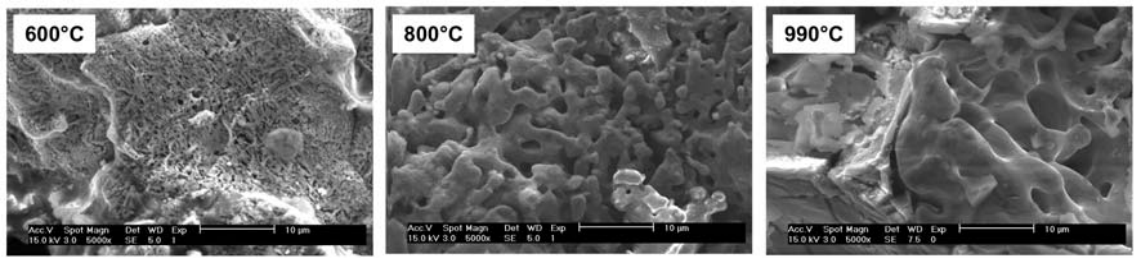

**Fig. 3.** Influence of the temperature on the morphology of the iron obtained after reduction of hematite cubes (5 mm side, 550 mg weight) by $H_2$ (200 mL/min of total gas flow, $H_2$/He 60/40 %vol.).

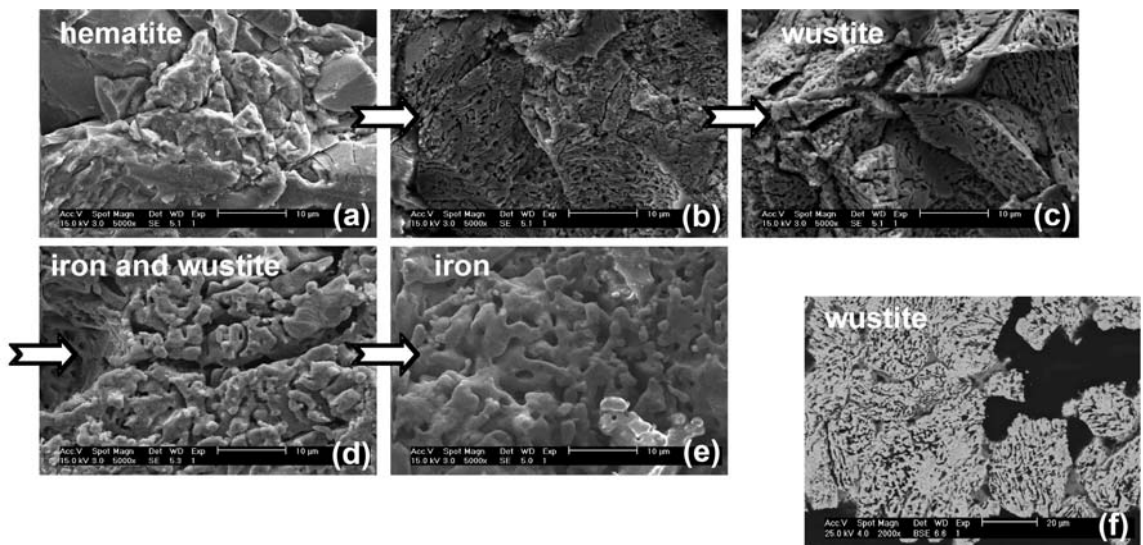

**Fig. 4.** Microstructure evolution of the solids during reduction at 800 °C by $H_2$ (200 mL/min of total gas flow, $H_2$/He 60/40 %vol.). Interrupted experiments: (a) initial; (b): after 89 s; (c): after 206 s; (d): after 400 s; (e): fully reduced; (f): after 400s, cross-section of polished sample

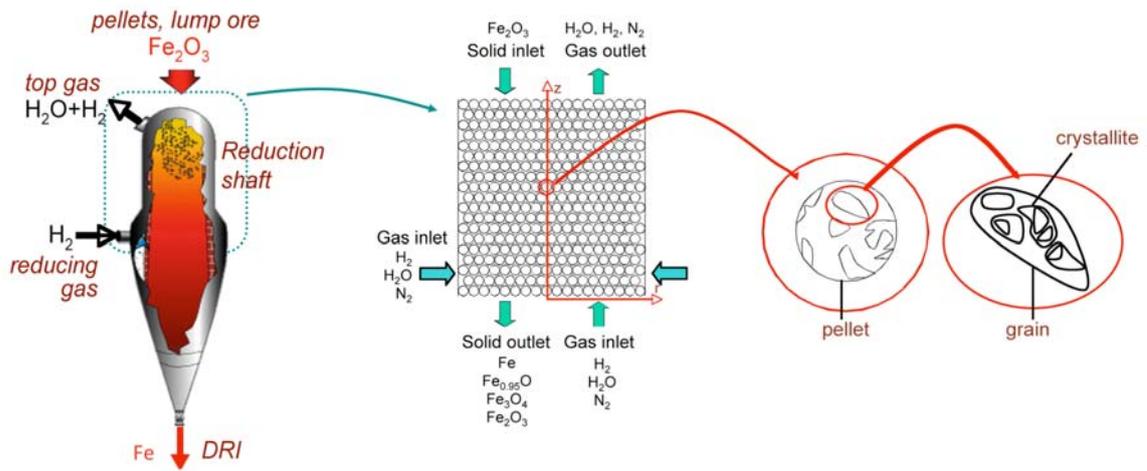

**Fig. 5.** Principle of the REDUCTOR model

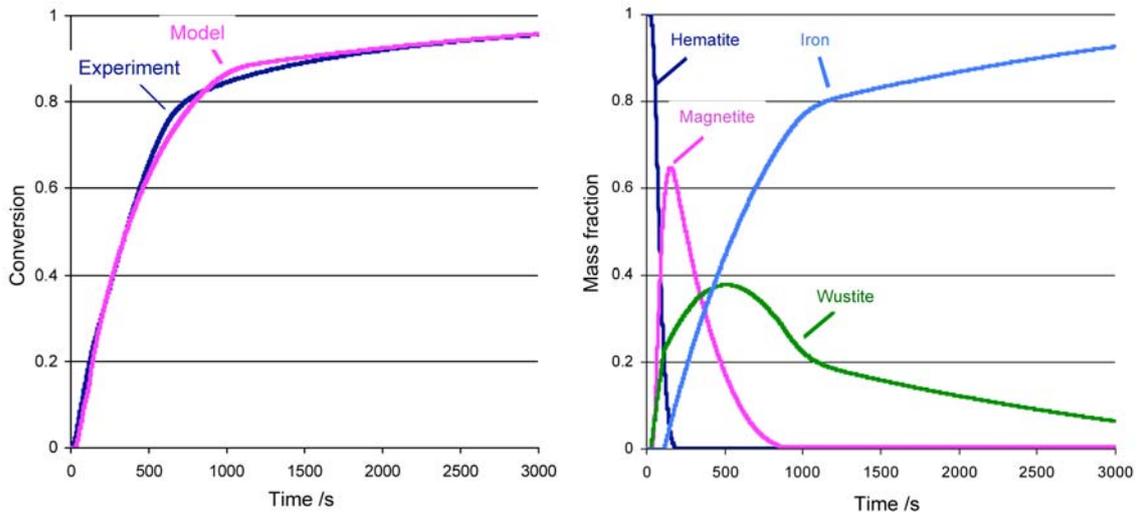

**Fig. 6.** Simulation results from the kinetic model describing the $H_2$ reduction of a single pellet at 900°C (200 mL/min of total gas flow $H_2$/He 60/40 %vol.): Left - comparison with thermogravimetric experiments; Right – evolution of the mass fractions of each oxide with time

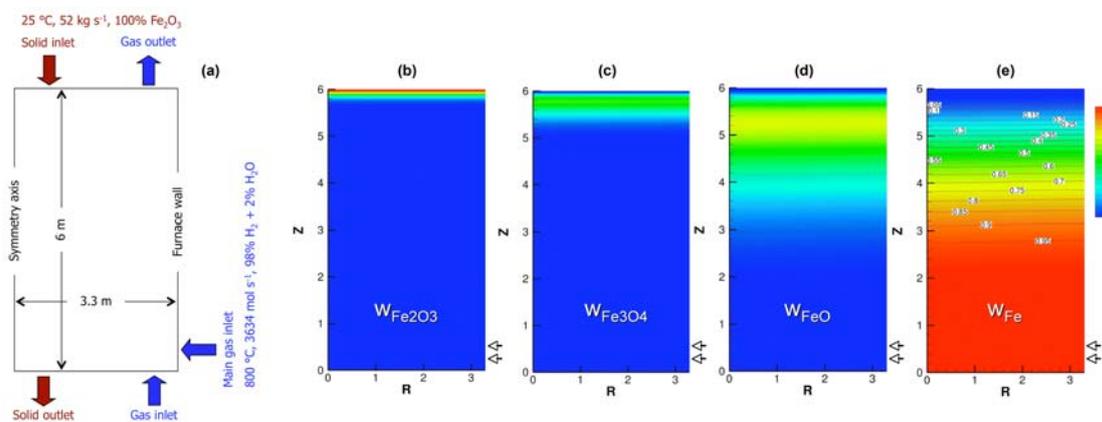

**Fig. 7.** Reference case: (a) operating parameters; (b), (c), (d), (e): calculated mass fraction of solid species inside the reactor.

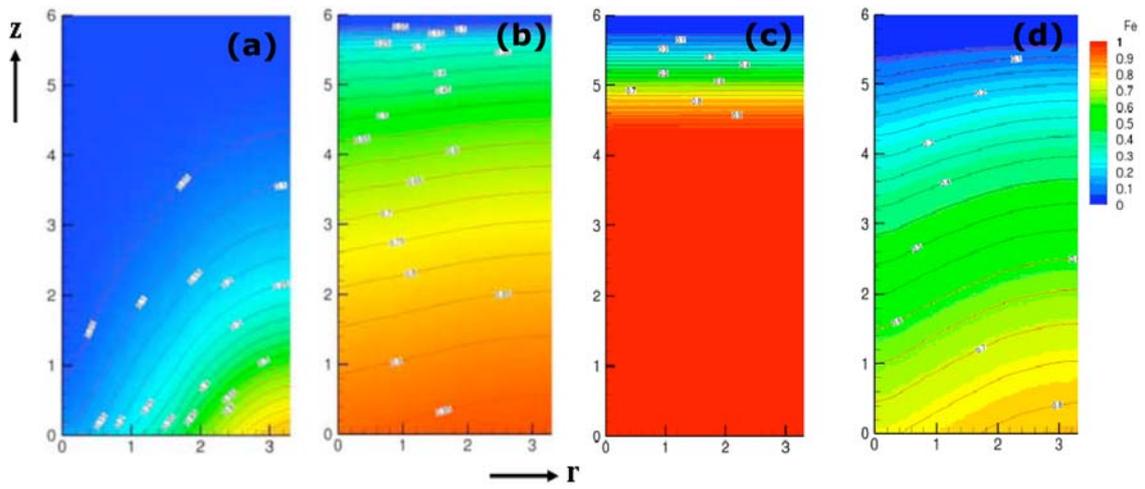

**Fig. 8.** Iron mass fraction inside the reactor: (a) $T_g$=600 °C; (b) $T_g$= 1000 °C; (c) $d_p$=6 mm; (d) $d_p$=24 mm